\documentstyle[epsfig,11pt]{article}

\newlength{\dinwidth}
\newlength{\dinmargin}
\newlength{\extraspace}
\newlength{\extraspaces}
\newcommand{\be}{\begin{equation}
\addtolength{\abovedisplayskip}{\extraspaces}
\addtolength{\belowdisplayskip}{\extraspaces}
\addtolength{\abovedisplayshortskip}{\extraspace}
\addtolength{\belowdisplayshortskip}{\extraspace}}
\newcommand{\ee}{\end{equation}}
\newcommand{\bdm}{\begin{displaymath}
\addtolength{\abovedisplayskip}{\extraspaces}
\addtolength{\belowdisplayskip}{\extraspaces}
\addtolength{\abovedisplayshortskip}{\extraspace}
\addtolength{\belowdisplayshortskip}{\extraspace}}
\newcommand{\edm}{\end{displaymath}}
\def\simlt{\mathrel{\lower2.5pt\vbox{\lineskip=0pt\baselineskip=0pt
           \hbox{$<$}\hbox{$\sim$}}}}
\def\simgt{\mathrel{\lower2.5pt\vbox{\lineskip=0pt\baselineskip=0pt
           \hbox{$>$}\hbox{$\sim$}}}}
\newcommand{\ls}[1]
   {\dimen0=\fontdimen6\the\font
    \lineskip=#1\dimen0
    \advance\lineskip.5\fontdimen5\the\font
    \advance\lineskip-\dimen0
    \lineskiplimit=.9\lineskip
    \baselineskip=\lineskip
    \advance\baselineskip\dimen0
    \normallineskip\lineskip
    \normallineskiplimit\lineskiplimit
    \normalbaselineskip\baselineskip
    \ignorespaces}

\catcode`@=11
\newcount\@tempcntc
\def\@citex[#1]#2{\if@filesw\immediate\write\@auxout{\string\citation{#2}}\fi
  \@tempcnta\z@\@tempcntb\m@ne\def\@citea{}\@cite{\@for\@citeb:=#2\do
    {\@ifundefined
       {b@\@citeb}{\@citeo\@tempcntb\m@ne\@citea\def\@citea{,}{\bf ?}\@warning
       {Citation `\@citeb' on page \thepage \space undefined}}%
    {\setbox\z@\hbox{\global\@tempcntc0\csname b@\@citeb\endcsname\relax}%
     \ifnum\@tempcntc=\z@ \@citeo\@tempcntb\m@ne
       \@citea\def\@citea{,}\hbox{\csname b@\@citeb\endcsname}%
     \else
      \advance\@tempcntb\@ne
      \ifnum\@tempcntb=\@tempcntc
      \else\advance\@tempcntb\m@ne\@citeo
      \@tempcnta\@tempcntc\@tempcntb\@tempcntc\fi\fi}}\@citeo}{#1}}
\def\@citeo{\ifnum\@tempcnta>\@tempcntb\else\@citea\def\@citea{,}%
  \ifnum\@tempcnta=\@tempcntb\the\@tempcnta\else
   {\advance\@tempcnta\@ne\ifnum\@tempcnta=\@tempcntb \else \def\@citea{--}\fi
    \advance\@tempcnta\m@ne\the\@tempcnta\@citea\the\@tempcntb}\fi\fi}
\catcode`@=12

\begin{document}

\setcounter{footnote}{1}
\begin{flushright}
LBNL-49221
\end{flushright}
\begin{center}
\Large{\bf     
      Rare $B$ Decays Beyond $B\to X_s\gamma$\footnote{
Invited talk given at the 9th Heavy Flavors conference, Caltech, Pasadena, 
Sept.10-13 2001.}}
\end{center}
\vspace{5mm}
\begin{center}
Gustavo Burdman\\
\vspace{0.3cm}
{\normalsize\it Theoretical Physics Group, Lawrence Berkeley 
National Laboratory,} \\ 
{\normalsize \it Berkeley, CA 94720 }
\end{center}

%


\begin{abstract}
\noindent
I discuss recent progress in our understanding of exclusive rare and 
semileptonic decays. I show the impact of HQET when 
combined with the predictions in the Large Energy Limit of QCD, 
focusing first on applications to $B\to K^*\ell^+\ell^+$. I 
also discuss the constraints
on semileptonic form-factors that appear in HQET/LEET with the use of 
radiative decay data, and 
update these to include the effects of next-to-leading order 
contributions in $B\to K^*\gamma$, as well as the latest data.    
\end{abstract}

\section{Introduction}

In the Standard Model (SM), Flavor Changing Neutral Currents (FCNCs) are 
forbidden at tree level.
They can occur starting at one loop. For the $b\to s$ transitions, the
diagrams involving the top quark dominate the short distance rate. This is 
essentially a consequence
of non-decoupling in spontaneously broken gauge theories: heavy fermions in 
the loops give contributions
that do not vanish as the mass increases but rather grow. Thus, FCNC 
processes at relatively
low energies have the potential to explore high energy scales such as the 
weak scale or even beyond.
Extensions of the SM, such as supersymmetry, technicolor, etc. would result 
in new contributions
to FCNC loops, leading to deviations from the SM predictions. This 
sensitivity of FCNC processes
to high energy scales, and therefore to new physics makes them of great 
interest. This is particularly
true in $B$ decays since it is possible that the third generation is 
involved in electroweak symmetry breaking
\footnote{For instance, this is the case in supersymmetry, where the 
stop plays a crucial role;
as well as in topcolor models where top condensation is responsible 
for (at least partially)
breaking electroweak symmetry.}.
However,~$B$ decays are affected by hadronic uncertainties that may 
obscure the interesting short
distance processes. Such uncertainties result from the fact that 
hadronization is a non-perturbative
problem only tractable from first principles by lattice gauge theory.
The use of inclusive decay modes greatly circumvent this problems. 
There, the uncertainties
are mostly not from hadronization (only relevant in the initial state) 
but from perturbative QCD.
Exclusive modes, on the other hand, are largely affected by theoretical 
uncertainties from 
hadronic matrix elements of the short distance operators.

While lattice calculations progress toward greater precision and accuracy, 
we can ask the question:
can rare $B$ decays be used as tests of the one-loop structure of the SM 
with our current
knowledge of hadronic matrix elements ? The comparison with the program of 
electroweak precision measurements, 
mostly performed at the $Z$ pole, is interesting.  That program relied on 
processes with large tree-level SM
amplitudes (large SM background), but exquisite experimental precision 
made it possible to achieve sensitivity
to one loop contributions since the accuracy of our theoretical knowledge 
matches this precision.
On the other hand, rare $B$ decays and FCNC processes in general, start 
at one loop
but are affected by large theoretical uncertainties.
These considerations are still valid even if we are talking about new 
physics entering at tree level. 
Very large data samples of $b\to s\gamma$ 
and $b\to s\ell^+\ell^-$ 
decays will be available soon. How good a test of the SM will these be? 
The inclusive $B\to X_s\gamma$ decay already 
constrains physics beyond the SM. Even more constraints will result once 
$b\to s\ell^+\ell^-$ modes begin to 
be measured. 

Inclusive modes are theoretically cleaner. The uncertainties are, in 
principle, from perturbative QCD and the 
OPE for heavy quarks. In practice, there is the added problem that cuts 
are needed to make contact with 
experiment~\cite{inc}. 
This introduces additional uncertainties. In addition, the 
experimental signals require a 
very clean environment. 
Exclusive decays are easier experimentally. But, as e mentioned earlier, 
they are affected by large hadronic
uncertainties coming as unknown form-factors. These are the kinds of 
things the lattice will one day compute
precisely. But for now (short of using models) 
we have to rely on symmetries and other related tricks in order to extract the
short distance physics from these modes. 

The factorization of short and long distance physics takes place in the 
effective hamiltonian at low energies. 
This is obtained by integrating out the heavy fields (e.g. the $W$, the 
top quark, the gluino, etc.)
and has the form 
\begin{equation}
{\cal H}_{eff}=-\frac{4G_F}{\sqrt{2}}V_{tb}^*V_{ts}
\sum_{i=1}^{10} C_i(\mu)\;   O_i(\mu) 
\label{heff}
\end{equation}
Here, the  basis $\{O_i\}$ constitute a complete set of operators 
leading to $b\to s$ transitions, 
whereas the Wilson coefficient functions $C_i(\mu)$ encode the short 
distance information coming from integrating
out the heavy degrees of freedom. For instance in the SM the $C_i(\mu)$ come 
from loop integrals involving 
the $W$ and the top quark and they will depend on their masses. If physics 
beyond the SM contributes to these decays
it will shift the values of the Wilson coefficients with terms now 
depending on gluino, squark or technipion masses.
The $\mu$ dependence of the Wilson coefficients is in principle canceled 
by that of  the matrix elements of the operators.
At leading order in $\alpha_s$ the operators mediating $b\to s$ transitions
are 
\begin{eqnarray}
{\cal O}_7&=&\frac{e}{16\pi^2}m_b \bar s_L\sigma_{\mu\nu} b_R \;F^{\mu\nu}
\label{o7}\\
{\cal O}_9&=&\frac{e^2}{16\pi^2}\;(\bar s_L\gamma_\mu b_L)
(\bar\ell\gamma^\mu\ell)
\label{o9}\\
{\cal O}_{10}&=&\frac{e^2}{16\pi^2}\;(\bar s_L\gamma_\mu b_L)
(\bar\ell\gamma^\mu\gamma_5\ell)\label{o10}~.
\end{eqnarray}
Among these, only ${\cal O}_7$ contributes to processes with the photon 
on-shell, such as $b\to s\gamma$. 
Mixing with other operators occurs due to the strong interactions.  
Most notably with the gluonic dipole operator ${\cal O}_8=(g/16\pi^2)
m_b \bar s_LT^a\sigma_{\mu\nu} b_R G^{a\mu\nu}$, and especially with 
$O_2=(\bar s_L\gamma_\mu c_L)(\bar c_L\gamma^\mu b_L)$, 
which is generated by tree-level $W$ exchange.


\section{Exclusive Decays}
The hadronic matrix elements can be parametrized in terms of form-factors. 
For the $B\to K\ell^+\ell^-$ decay 
the hadronic matrix elements of the operators $O_7$, $O_9$ and $O_{10}$ 
can be written as 
\begin{eqnarray}
\langle K(k)|\bar{s}\sigma_{\mu\nu}q^\nu b|B(p)\rangle &=& i\frac{f_T}{m_B+m_K}
\left\{(p+k)_\mu q^2-  q_\mu(m_B^2-m_K^2)\right\}~, \label{bp_sig} \\
\langle K(k)|\bar{s}\gamma_\mu b|B(p)\rangle &=& f_+ (p+k)_\mu +f_- q_\mu ~,
\label{bp_sem}
\end{eqnarray}
with $f_T(q^2)$, and $f_{\pm}(q^2)$ unknown functions of $q^2=(p-k)^2=
m_{\ell^+\ell^-}^2$. 
 In the SU(3) limit $f_{\pm}$ in (\ref{bp_sem}) are the same as the 
form-factors entering in the semileptonic decay $B\to\pi\ell\nu$. 
For $B \to K^* \ell^+\ell^-$ decays of the
``semileptonic'' matrix elements over vector and axial vector currents are
\begin{eqnarray}
\langle K^*(k,\epsilon)|\bar{q}\gamma_\mu b|B(p)\rangle
&=& \frac{2V(q^2)}{m_B+m_V}\;\epsilon_{\mu\nu\alpha\beta}
\epsilon^{*\nu}p^\alpha k^\beta 
\label{vcurr}\\
\langle K^*(k,\epsilon)|\bar{q}\gamma_\mu\gamma_5 b|B(p)\rangle
&=& i 2m_V\,A_0(q^2)\frac{\epsilon^*\cdot q}{q^2}q_\mu
+ i (m_B+m_V)\,A_1(q^2)\,\left(\epsilon^*_\mu - 
\frac{\epsilon^*\cdot q}{q^2}q_\mu
\right) \nonumber\\
& &-i A_2(q^2)\,\frac{\epsilon^*\cdot q}{m_B+m_V}\left( (p+k)_\mu 
-\frac{m_B^2-m_V^2}{q^2}q_\mu\right)~.
\label{vacur}
\end{eqnarray}
and for the FCNC magnetic dipole operator $\sigma_{\mu \nu}$ 
\begin{eqnarray}
\langle K^*(k,\epsilon)|\bar{q}\sigma_{\mu\nu}(1+\gamma_5) q^\nu b|B(p)\rangle &
=& i\, 2T_1(q^2) 
\epsilon_{\mu\nu\alpha\beta}\epsilon^{*\nu}p^\alpha k^\beta\nonumber\\
& &+ T_2(q^2)\, \left\{ \epsilon_\mu^*(m_B^2-m_{V}^2) - 
(\epsilon^*\cdot p)\,(p+k)_\mu \right\}\nonumber \\
& &+T_3(q^2) (\epsilon^*\cdot p)\,\left\{q_\mu 
-\frac{q^2}{m_B^2-m_{V}^2}\, (p+k)_\mu \right\}~, \label{sig2kst}
\end{eqnarray} 
where $\epsilon_\mu$ denotes the polarization four-vector of the
$K^*$. 
In the $SU(3)$ limit these form-factors also describe
$B\to\rho\ell\nu$, as well as $B_s\to\phi\ell^+\ell^-$ decays.
The form-factors defined in eqns.(\ref{bp_sem})-(\ref{sig2kst}) 
are not calculable in a controlled approximation with the notable
exception of lattice gauge theory. 
We will see however, that symmetries and generally the behavior of 
hadrons in certain limits will allow us to gain important insight on 
these quantities.
Notice that $T_1(0)=T_2(0)$ and $T_3$ does not contribute to the
amplitude to the radiative decay into an on-shell photon, i.e. in 
$B\to K^*\gamma$. So now $Br(B\to K^*\gamma)~\alpha~
\,|T_1(0)|^2$, where the short distance is basically the same as in inclusive
decays\footnote{With some caveats and exceptions we discuss later on. See  
also~\cite{stefan}.}.

\subsection{Heavy Quark Symmetry and $B\to K^*\ell^+\ell^-$}
In the HQL~\cite{iwsym} $m_b\gg \Lambda_{\rm QCD}$ the form-factors 
${T_i(q^2)}$ corresponding to the dipole operator are not independent 
of the semileptonic form-factors $V(q^2),{A_i(q^2)}$.  
Instead, they obey the following relations 
\cite{iw90,bd91}
\begin{eqnarray}
  T_1(q^2) &=& \frac{m_B^2+q^2-m_V^2}{2m_B}\frac{V(q^2)}
   {m_B+m_V}+\frac{m_B+m_V}{2m_B} A_1(q^2),
\nonumber\\
\hspace*{-0.5cm}\frac{m_B^2-m_V^2}{q^2}\Big[T_1(q^2)-T_2(q^2)\Big] &=&
   \frac{3m_B^2-q^2+m_V^2}{2m_B}\frac{V(q^2)}
   {m_B+m_V}-\frac{m_B+m_V}{2m_B} A_1(q^2),
\label{rel_3}\\
  T_3(q^2) &=& \frac{m_B^2-q^2+3m_V^2}{2m_B} \frac{V(q^2)}{m_B+m_V}+
\frac{m_B^2-m_V^2}{m_B q^2} m_V A_0(q^2) 
\nonumber\\  &\hspace*{-1cm}-&\hspace*{-0.5cm}\frac{m_B^2+q^2-m_V^2}{2m_B q^2}
  \Big[(m_B+m_V)A_1(q^2)-(m_B-m_V)A_2(q^2)\Big].
\nonumber 
\end{eqnarray}
In terms of the symmetries of the HQET, eqns.~(\ref{rel_3}) 
result from the Heavy Quark {\em Spin} Symmetry (HQSS)
that arises in the 
heavy quark limit due to the decoupling of the spin of the heavy 
quark~\cite{iwsym}. 
These relations imply, for instance, that the 
$B\to\rho\ell\nu$ and $B\to\rho\ell^+\ell^-$ are described by the same 
form-factors. Additionally, if $SU(3)$ symmetry is assumed, these also 
are the $B\to K^*\ell^+\ell^-$ form-factors.

\subsection{The Large Energy Limit}
We now consider the Large Energy Limit (LEL)~\cite{lel1} for heavy-to-light
transitions into a vector meson as the ones we are studying. 
As a result, one recovers the HQSS 
form-factor relations (\ref{rel_3}), 
but now there are additional new relations among the form-factors defined in 
(\ref{vcurr}-\ref{sig2kst}). 
These read as~\cite{lel2}
\begin{eqnarray}
V(q^2) &=& \left(1+\frac{m_V}{M}\right)\,\xi_{\perp}(M,E)~,\label{le4}\\
A_1(q^2) &=& \frac{2E}{M+m_V} \,\xi_{\perp}(M,E)~,\label{le5}\\
A_2(q^2) &=&\left(1+\frac{m_V}{M}\right)\,\left\{
\xi_\perp(M,E) - \frac{m_V}{E}\xi_\parallel(M,E)\right\}~,
\label{le6}\\
A_0 (q^2)&=& \left(1-\frac{m_V^2}{ME}\right)
\xi_\parallel(M,E) + \frac{m_V}{M}\xi_\perp(M,E)~,
\label{le7}
\end{eqnarray}
and
\begin{eqnarray}
T_1(q^2) &=& \xi_\perp(M,E)~,\label{t1}\\
T_2(q^2) &=& \left(1-\frac{q^2}{M^2-m_V^2}\right) \xi_\perp(M,E)~,
\label{t2}~,\\
T_3(q^2)&=&\xi_\perp(M,E) - 
\frac{m_V}{E}\left(1-\frac{m_V^2}{M^2}\right)\xi_\parallel(M,E)~.
\label{t3}
\end{eqnarray}
and will receive corrections that roughly go as 
$(\Lambda_{QCD})/E_h$.
It is apparent from eqns.~(\ref{le4})-(\ref{t3}) that, in the LEL regime, 
the $B\to V\ell\ell'$ decays are described by only two form-factors:
$\xi_\perp$ and $\xi_\parallel$, instead of the seven apriori independent
functions in the general Lorentz invariant ansatz of the matrix
elements. Here, 
$\xi_\perp$ and $\xi_\parallel$ are functions of the heavy mass
$M$ and the hadronic energy $E$, and refer to the transverse and longitudinal
polarizations, respectively.

This simplification leads to new relations among the form-factors. 
For instance, the ratio of the vector form-factor $V$
to the axial-vector form-factor $A_1$, 
\begin{equation}
R_V(q^2)\equiv\frac{V(q^2)}{A_1(q^2)}=\frac{(m_B+m_V)^2}{2E_Vm_B}~,
\label{rv}
\end{equation}
is independent of any of these unknown, non-perturbative functions
$\xi_{\perp,\parallel}$
and is determined by
purely kinematical factors. Here, $E_V=(m_B^2+m_V^2-q^2)/(2 m_B)$ denotes
the energy of the final light
vector meson. A similar relation holds 
for $T_1$ and $T_2$, since they both are also proportional to the  
``transverse'' form-factor $\xi_\perp$. 
As we will see below, these predictions have important consequences 
for observables at large recoil energies (low $q^2$). 

\subsection{Corrections to LEET}
The predictions obtained in the LEL receive corrections from several sources. 
In the past year or so some of them have been extensively 
addressed in the literature. Here is a quick review.

\flushleft\underline{\em Hard Corrections:} 
These corrections result from the exchange of hard gluons. 
There are two kinds of them: factorizable and non-factorizable 
gluon exchange. 
The first kind corresponds to either the
renormalization of the heavy-light currents which already appears 
in Ref.~\cite{iwsym}, or  
hard gluon exchange with the spectator quark.  The latter 
can be computed in the Brodsky-Lepage formalism for 
exclusive processes at large momentum transfers~\cite{bl}. 
This was done in Ref.~\cite{beneke1}, where it was found that the form-factor
relations in (\ref{le4})-(\ref{t3}) receive $\alpha_s$ corrections that
are typically $10\%$ or smaller.
An interesting  case is the ratio of vector to axial-vector
form-factors in eqn.~(\ref{rv}), which receives 
{\em no $\alpha_s$ corrections} 
to leading order in the $1/E$ expansion. This has an interesting explanation
and we will come back to this point below. 

The second kind of hard gluon corrections are the so called non-factorizable
ones~\cite{stefan,bfs}. 
These are mediated by diagrams where the gluon exchanged 
with the spectator comes from either the insertion of the operator 
${\cal O}_8$, or from the insertion of four-fermion operators 
${\cal O}_1-{\cal O}_6$ with the gluon attached to the quark loop.
These corrections cannot be absorbed by form-factors or renormalization of the 
currents. Thus they are genuinely distinct hard corrections that do not 
occur in inclusive decays. In Refs.~\cite{stefan,bfs} 
the effects of all these contributions are computed and are rather
large in $B\to K^*\gamma$. The effective ``exclusive'' Wilson coefficient
$C^{\rm excl.}_7$ is shifted by roughly $30\%$, whereas the effect in 
$C_9$ is smaller. The authors go on to compute the exclusive rate by making use
of light-cone QCD sum rules for the form-factors, resulting in a rate which is 
approximately a factor of two larger than the experimental rate. 
This may be signaling a problem with the form-factor calculation rather than 
one with the non-factorizable hard gluons. 

\flushleft\underline{\em Collinear Gluons:} 
In Ref.~\cite{bauer} an effective field theory including 
collinear quarks and gluons is developed. This theory includes
LEET but, unlike LEET,  it has the correct infrared behavior. 
Sudakov logarithms are accounted for in this treatment. 
Concerning its application to exclusive decays, this ``complete''
LEET does not modify the form-factor relations implied by 
eqns.(\ref{le4}-\ref{t3}). A main ingredient of LEET is preserved when the 
interaction with collinear gluons is taken into account: 
energetic quarks still are two component spinors in the LEL. 
Thus the complete LEET including collinear gluons confirms that
the LEET relations are valid in the large energy limit of QCD.

\flushleft\underline{\em Power Corrections:} 
Just as in HQET the $1/m_Q$ corrections are potentially a very 
important source of
uncertainty, the LEL predictions are affected by $1/E$ corrections. 
Unlike in HQET, it is not clear that LEET is a framework where the 
$1/E$ corrections can be estimated. Even when collinear gluons 
are incorporated, it is difficult to separate degrees of freedom
to be integrated out so that an effective field theory can emerge. 
Perhaps, we can still estimate the size of these corrections with this 
caveat in mind. More work is needed, and certainly once experimental 
tests of LEET predictions become available we hope to understand more about
them.

\subsection{Forward-Backward Lepton Asymmetry}
The forward-backward asymmetry for leptons as a function of the
dilepton mass squared $m_{\ell\ell}^2=q^2$ is defined as
\begin{equation}
A_{FB}(q^2)=\frac{
\int_{0}^{1}\frac{d^2\Gamma}{dxdq^2} dx - 
\int_{-1}^{0}\frac{d^2\Gamma}{dxdq^2} dx 
}
{\frac{d\Gamma}{dq^2}}  ~,
\label{afbdef}
\end{equation}
where $x\equiv\cos\theta$, and $\theta$ is the angle between the 
$\ell^-$ and the $\bar{B}^0$ in the dilepton center-of-mass frame
\footnote{In Ref.~\cite{gbafb} it is erroneously stated that $\theta$ 
is defined with respect to $\ell^+$. But the expressions there correspond to 
the current definition. This explains the sign difference with respect to 
Ref.~\cite{bfs}.}. 

The asymmetry is proportional to the Wilson coefficient $C_{10}$ 
and vanishes with it. 
Furthermore, it is proportional to a combination of $C_9^{\rm eff.}$ and 
$C_7^{\rm eff.}$ such that it 
has a zero in the physical region 
if the following condition is satisfied~\cite{gbafb}
\begin{equation}
Re[C_9^{\rm eff.}]=\frac{m_b}{q^2_0}\,C_7^{\rm eff.}\,\left\{
\frac{T_1}{V}(m_B+m_{K^*}) - 
(m_B-m_{K^*})\,\frac{T_2}{A_1}\right\}~,\label{afbz}
\end{equation}
where $q^2_0$ is the position of the $A_{FB}$ zero and all $q^2$-dependent 
quantities\footnote{Note that
the sign difference in eqn.(\ref{afbz}) with respect to a similar expression 
in Ref.\cite{ali} is due to a sign in the definition of $V$.}
are evaluated at $q^2_0$. 
In inclusive $B\to X_s\ell^+\ell^-$, 
the zero of the asymmetry implies the relation: 
$Re[C_9^{\rm eff.}]=-2 (m_B^2/q_0^2)\,C_7^{\rm eff.}$. Thus, in principle the 
condition in (\ref{afbz}) is affected by the presence of the hadronic 
form-factors
making it a priori a more uncertain relation. 
However, by making use of the HQSS relations (\ref{rel_3})
for $T_1(q^2)$ and $T_2(q^2)$, the form-factor eqn.~(\ref{afbz}) simplifies
to
\begin{equation}
\frac{Re[C_9^{\rm eff.}]}{ C_7^{\rm eff.}}=-\frac{m_b}{q^2_0}\left\{
\frac{2m_B\,k^2}{(m_B+m_{K^*})^2}R_V
+\frac{(m_B+m_{K^*})^2}{2m_BR_V}
+ 2(m_B-E_{K^*})
\right\}
\label{con2}
\end{equation}
where $R_V$ is the ratio of vector to axial-vector form-factors
defined in (\ref{rv}) and evaluated at $q_0^2$. Thus, the determination
of the zero of $A_{FB}(q^2)$ in $B\to K^*\ell^+\ell^-$
gives a relation between the short distance 
Wilson coefficients $C_9^{\rm eff.}$ and $C_7^{\rm eff.}$ where the only 
uncertainty
from form-factors is in the ratio $R_V$. 
As mentioned earlier, since (\ref{con2}) was derived using the heavy quark
{\em spin} symmetry, it is expected to receive small corrections. 
In principle, information on $R_V(q^2)$ could be extracted from 
the $B\to\rho\ell\nu$ decay. But it turns out that this may not be necessary.
If we plot the asymmetry vs. the dilepton mass in a variety
of models we note that the asymmetry is agreed upon with exceptional
accuracy! In Ref.\cite{gbafb} it was noted that 
\begin{figure}
\begin{center}
\includegraphics[height=.4\textheight,angle=90]{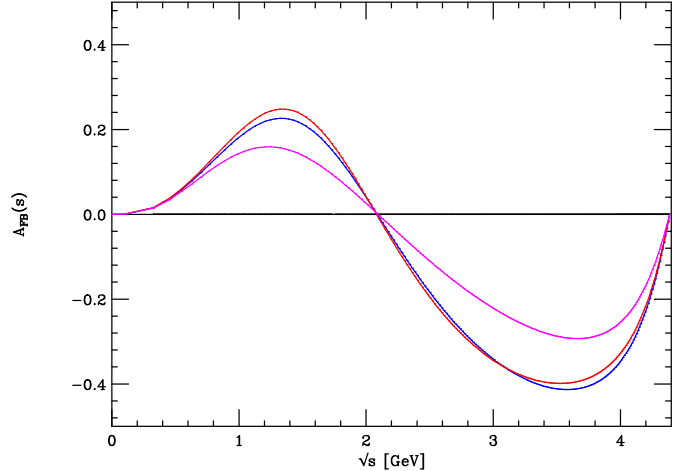}
  \caption{Differential forward-backward asymmetry vs. the dilepton mass 
squared for various model calculations.This includes all hard corrections, 
both factorizable and non-factorizable, so it 
is an update of Ref.~\cite{gbafb}, where the references for the models can be 
obtained.
}
\label{afb}
\end{center}
\end{figure}
this feature must emerge from a ``factorization'' of the 
soft physics in such a way that 
\begin{equation}
{ V(q^2)}\simeq A_{\rm kin.}\times {F_{\rm soft}},~~~~~~~~~~
{A_1(q^2)}\simeq B_{\rm kin.}\times {F_{\rm soft}}~,\nonumber
\end{equation} 
so that the soft physics cancels in the ratio $R_V$.
It was first recognized in Ref.~\cite{ali} that 
this is precisely what the LEET predicts: 
$R_V(q^2)=(m_B+m_{K^*})^2/(2m_BE_{K^*})$ 
as extracted
from eqns.~(\ref{le4}) and (\ref{le5}). Then, the condition for the
vanishing of $A_{FB}(q^2)$ reads now
\begin{equation}
Re[C_9^{\rm eff.}]=-2\frac{m_bm_B}{q_0^2}\;C_7^{\rm eff.}\,
\left(1-\frac{m_{K^*}^2}{2m_BE}\right)~.
\label{condlet}
\end{equation}
Thus, in the LEL the position of the zero of $A_{FB}(q^2)$
in $B\to K^*\ell^+\ell^-$ is predicted
in terms of the short distance Wilson coefficients $C_9^{\rm eff.}$ and 
$C_7^{\rm eff.}$. 
The hard gluon corrections discussed above would shift the position
of the zero by a calculable amount given in Ref.\cite{bfs}. There it is
pointed out that non-factorizable contributions change the value
of $q^2_0$ 
from what it would be obtained if the ``inclusive'' Wilson coefficients
are used by an amount around $(20-30)\%$. This is roughly $q_0^2=(4.2\pm0.6)
~{\rm GeV}^2$. The plot in Fig.\ref{afb} takes all these corrections into 
account.

\subsection{Semileptonic Form-factors in the Large Energy Limit}
We finally discuss another application of LEET. 
In Ref.~\cite{gg} the LEET prediction for $R_V$ in eqn.(\ref{rv}) 
was used in combination 
with the $B\to K^*\gamma$ and $B\to X_s\gamma$ data, 
and Heavy Quark Spin Symmetry
in order to constrain the semileptonic form-factors $V$ and $A_1$ at 
$q^2=0$. Here we update this analysis by incorporating the 
factorizable and non-factorizable hard corrections from 
Refs.\cite{stefan,bfs}.

\begin{figure}
\begin{center}
\includegraphics[height=.4\textheight,angle=90]{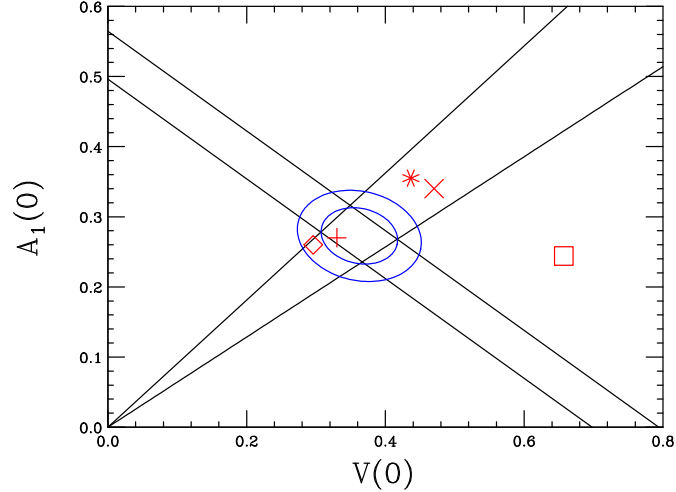}
 \caption{Constraints on the semileptonic form-factors $V(0)$ and 
$A_1(0)$ from 
$B^0\to K^{*0}\gamma$ data plus HQSS (thicker band) together with the
relation from the LEL (cone).The ellipses correspond to $68\%$ and $95\%$ 
confidence level intervals. 
Central values of model predictions are also shown and correspond to 
BSW~\cite{bsw} (vertical cross), ISGW2~\cite{isgw2} (diamond),
 MS~\cite{ms} (star), LCSR~\cite{ali} (diagonal cross) and
LW~\cite{lw} (square), respectively. This updates Ref.~\cite{gg} to include
hard corrections as computed in Ref.~\cite{stefan,bfs}, as well as 
the most recent data.}
\label{fitcon}
\end{center}
\end{figure}

Since the NLO corrections in the exclusive mode are not canceled by the
ones in the inclusive decay, we will not normalize the exclusive branching 
ratio to the inclusive one. Instead we make use of the exclusive 
data only in order to extract the $B\to K^*\gamma$ form-factor. 
The branching ratio reads
\begin{equation}
Br(B\to K^*\gamma)
= \tau_B\,\frac{\alpha\,G_F^2}{32\pi^4}\,|V_{cb}V^*_{cs}|^2\,
m_b^2\,m_B^3\,(1- \frac{m_{K^*}^2}{m_B^2} )^3 
\,\left|A_7\right|^2 \,|T_1(0)|^2~,
\label{rgamma}
\end{equation}
where the form-factor $T_1(q^2)$ was defined in (\ref{sig2kst}), and 
$A_7$ is the NLO effective Wilson coefficient for the exclusive mode,  
which differs from the one entering 
in inclusive decays. 
We use the running mass $m_b(m_b)=(4.2\pm0.2)~$GeV, 
and $|V_{cb}V_{cs}^*|=0.04$. 
The value of $A_7$ was computed in 
Ref.~\cite{stefan} to be $A_7=-0.4072 - i0.0256$. 
We make use of the world average for the neutral 
mode~\cite{bsgexp}
$Br(B^0\to K^{*0}\gamma)=(4.56\pm0.37)\times10^{-5}$. 
We obtain 
$T_1(0) =0.31\pm0.02$, where the error mainly reflects the experimental 
uncertainty, the 
uncertainty in the $b$ quark mass and the scale dependence 
in $A_7$. 
This result assumes the SM for the calculation of the 
short distance coefficient $A_7$. However, this is not a very strong assumption
given that the agreement of the inclusive calculation with 
the $B\to X_s\gamma$, and the fact that the new physics is likely  to affect
this rate and the exclusive one similarly.
Armed with this extracted value of $T_1(0)$, now we can go to the 
HQSS relations and turn them into a relation between 
$V(0)$ and $A_1(0)$. This is 
\begin{equation}
A_1(0)=\frac{2 m_B}{m_B+m_{K^*}} \, T_1(0)-\frac{m_B-m_{K^*}}{m_B+m_{K^*}}
\,V(0)~.
\label{expconst}
\end{equation} 
Eqn.~(\ref{expconst}) results in a constraint in the $(V(0),A_1(0))$ plane, 
which in Fig.~\ref{fitcon}
corresponds to the band descending from left to right. 
In addition, the expression (\ref{rv}) for $R_V$ gives
another constraint corresponding to a straight line going through the origin.
This LEET prediction will surely receive corrections at next to leading order
in the $1/E$ and $1/m_b$ expansions. These corrections have not been computed
and we simply guess they are of order $m_{K^*}/m_B\simeq0.17$, giving the cone 
in Fig.~\ref{fitcon}. We should notice, however, that the LEET prediction
for $R_V$ does not receive hard gluon or collinear gluon corrections. 
We will come back to this point. 
The ellipses in Fig.~\ref{fitcon} correspond to $68\%$ and $95\%$ C.L.
intervals. The fit gives 
\begin{equation}
V(0)=0.36\pm0.04~,~~~~~~~~~~~~~~~~~~~A_1(0)=0.27\pm0.03~. 
\label{result}
\end{equation}
The results in eqn.~(\ref{result}) differ from Ref.~\cite{gg} where the NLO
corrections had not been taken into account. 
These lift the value of the Wilson coefficient by roughly $30\%$
resulting in a lower value for $T_1(0)$, and through the fit in 
Fig.~\ref{fitcon}, in 
lower values for both $V(0)$ and $A_1(0)$. 
Also, here we use the most recent average of the neutral mode
resulting in a thinner band than the one in Ref.~\cite{gg}.

We compare our findings for $V(0)$ and $A_1(0)$ with
several model predictions. For illustration, we take 
the Bauer-Stech-Wirbel (BSW) model from Ref.~\cite{bsw}
(cross), the modified 
version of the Isgur-Scora-Grinstein-Wise (ISGW2) model from 
Ref.~\cite{isgw2} (diamond), 
a recent relativistic constituent
quark model prediction by  Melikhov and Stech (MS) \cite{ms} (star),
a recent calculation in the Light Cone QCD Sum Rule (LCSR) formalism
of Ref.~\cite{ali} (diagonal cross)  and 
the prediction by Ligeti and Wise (LW) from Ref.\cite{lw} (square).
As it can be seen from Fig.~\ref{fitcon}, the latter is even more excluded
now. As discussed in more detail in Ref.\cite{gg}, this is likely 
to be due  largely 
to the use of a monopole form-factor to extrapolate from the 
charm data to the maximum recoil in a $B$ decay. Corrections to the 
heavy quark flavor symmetry, although large, are unlikely to be the sole 
explanation for such a discrepancy.
A new finding in this update
is the apparent exclusion of  the MS and LCSR
predictions for $V(0)$ and $A_1(0)$. These models, were in agreement 
with the fit in Ref.~\cite{gg}.  
Of course, this exclusion refers to the predicted central values for the 
models. In some cases the predictions have large errors and modifications 
in the calculations may bring them into agreement with the fit. 

A potentially large isospin violation splitting the neutral and charged
modes was found in Ref.\cite{kn} in the context of QCD factorization. 
This would shift the value of $A_7$ in the neutral mode by a few
percent, resulting in a similarly small shift in the value of 
$T_1(0)$ extracted from eqn.(\ref{rgamma}).

Finally, we comment on the leading order
LEET prediction for $R_V$, eqn.~(\ref{rv}).  
Making use of the HQSS relations (\ref{rel_3}),
the transverse helicity amplitudes for a generic 
$B^-\to V^-\ell\ell'$ transition 
can be written as
\begin{equation}
 H_\pm = {\cal F} \;(V\mp\frac{(m_B+m_V)^2}{2m_Bk_V}A_1)~,
\label{hel}
\end{equation}
where $\cal F$ is a factor depending on the mode under consideration
(e.g.~Wilson coefficients, coupling constants, etc...)
and $k_V$ is the momentum of the vector meson. 
Thus, we see from the form of $R_V$ in the large energy limit, 
that the ``$+$'' helicity vanishes $H_+=0$ in the LEL regime, up to residual 
terms of order $m_V^2/2E_V^2$.
This is not a surprise: in the 
limit of an infinitely heavy quark decaying into a light quark, 
the helicity of the latter is ``inherited'' by the final vector meson.
In the SM, the $(V-A)$ structure in semileptonic decays is reflected
in the dominance of the $H_-$ transverse helicity. 
On the other hand, the amplitude to flip the helicity of the fast 
outgoing light quark is suppressed by $1/E_h$. 
Quark models tend to have this concept built in, which explains the agreement
on the position of the zero in Fig.~\ref{afb}.
This is also the reason why $\alpha_s$ corrections from hard
gluon exchange between the spectator quark and the fast light quark
do not affect eqn.~(\ref{rv}): they are not helicity-changing at leading 
order in $1/E_h$.  
The same is true for the ratio
of $T_1$ and $T_2$.

\vspace{0.4cm}
{\bf Acknowledgments}

\noindent
I thank Gudrun Hiller, with whom part of the work reported 
here was done, for discussions and a careful reading of the manuscript.
I also thank the organizers of HF9 at Caltech for
such an excellent meeting, in spite of  
the difficult circumstances they had to face during 
its course. This work was supported by the Director, Office of Science, 
Office of High Energy and Nuclear Physics of the U.S. Department of 
Energy under Contract DE-AC0376SF00098. 




\end{document}